\pdfoutput=1

\documentclass[aps,prl,twocolumn,superscriptaddress,showpacs]{revtex4}
\usepackage{graphics}
\usepackage{epstopdf}
\usepackage{graphicx}
\usepackage{dcolumn}
\usepackage{bm}
\usepackage{amsmath}

\usepackage{color}
\usepackage{cancel}

\newcommand{\comm}[1]{{}}

\newcommand{\be}{\begin{equation}}
\newcommand{\ee}{\end{equation}}
\newcommand{\beq}{\begin{eqnarray}}
\newcommand{\eeq}{\end{eqnarray}}

\def\nue{\mathrel{{\nu_e}}}
\def\numu{\mathrel{{\nu_\mu}}}

\def\barnue{\mathrel{{\bar \nu}_e}}
\def\barnumu{\mathrel{{\bar \nu}_\mu}}

\def\t13{\mathrel{{\theta_{13}}}}
\def\y12{\mathrel{{\tan^2 \theta_{12}}}}
\def\c2{\mathrel{{\chi^2 }}}

\newcommand{\n}{neutrino}
\newcommand{\ns}{neutrinos}

\newcommand{\fb}{{FB}}
\newcommand{\ic}{IceCube}
\newcommand{\nt}{KM3NeT}

\begin{document}


\title{High Energy Neutrinos from the Fermi Bubbles}

\author{Cecilia Lunardini}
 \email{Cecilia.Lunardini@asu.edu}   
\affiliation{Arizona State University, Tempe, AZ 85287-1504}%
\affiliation{RIKEN BNL Research Center, Brookhaven National Laboratory, Upton, NY 11973}

\author{Soebur Razzaque}
\email{srazzaqu@gmu.edu}
\affiliation{School of Physics, Astronomy and Computational Sciences, 
George Mason University, Fairfax, VA 22030,
resident at Naval Research Laboratory, Washington, DC 20375}

 
\begin{abstract}
Recently the {\em Fermi}-LAT data have revealed two gamma-ray emitting bubble-shaped structures at the Galactic center. If the observed gamma rays have hadronic origin (collisions of accelerated protons), the bubbles must emit high energy neutrinos as well.  This new, Galactic, neutrino flux should trace the gamma ray emission in spectrum and spatial extent.  Its highest energy part, above 20--50 TeV, is observable at a kilometer scale detector in the northern hemisphere, such as the planned KM3NeT, while interesting constraints on it could be obtained by the IceCube Neutrino Observatory at the South pole. The detection or exclusion of neutrinos from the {\em Fermi} bubbles will discriminate between hadronic and leptonic models, thus bringing unique information on the still mysterious origin of these objects and on the time scale of their formation.
\end{abstract}                            
 
\pacs{95.85.Ry, 14.60.Pq, 98.70.Sa}
\maketitle

Gamma-ray data in the 1--100 GeV range from the {\em Fermi}-LAT~\cite{Atwood:2009ez} show a new, unexpected, feature of our galaxy: two huge spheroidal structures ({\em Fermi} ``bubbles", FB from here on), extending up to $\sim 8-9$ kpc ($50^\circ$) out of the galactic center on either side of the galactic disk~\cite{Dobler:2009xz,Su:2010qj}.  Intriguingly, the bubbles coincide spatially with the  WMAP ``haze" in microwave~\cite{Finkbeiner:2003im} and the thermal X-ray emission seen by ROSAT~\cite{Snowden:1997ze}. The origins of these three phenomena, and whether they are related, are matters of intense debate~\cite{Dobler:2009xz,Su:2010qj,Malyshev:2010xc,Crocker:2010dg,Mertsch:2011es,Cheng:2011xd}.

If we limit ourselves to the Standard Model of particle physics, there are two main models for the \fb. One is the leptonic origin: Compton scattering of relativistic electrons on photons (microwave, or optical/ultraviolet), where the electrons originate from shocks in the  outflow of the Galactic center black hole, Sgr A$^*$~\cite{Su:2010qj,Mertsch:2011es}, or from an episodic activity of Sgr A$^*$ due to the capture of a star \cite{Cheng:2011xd} or to a localized star formation (SF) event \cite{Zubovas:2011py}.  The second scenario is the hadronic origin: collision of accelerated protons on background protons in the bubble gas ~\cite{Crocker:2010dg}.  The collisions produce $\pi^0$ mesons, which then decay as $\pi^0 \rightarrow \gamma \gamma$. Here the accelerated protons originate in  the supernova remnants  (SNRs) near the Galactic center.  The two scenarios differ  widely in the cooling times of the accelerated particles, and therefore in the predicted age of the bubbles: millions of years for the leptonic model, versus billions for the hadronic case. Thus, the origin of the \fb\ will teach us about the time scale of Sgr A$^*$'s activity and about the Galactic star-formation history. 

A major signature of the hadronic models -- and therefore a discriminator between the leptonic and hadronic models of the \fb\ \cite{Crocker:2010dg,Vissani:2011ea} -- is the presence of a high energy, $\sim 1$--10 TeV neutrino ($\nu$) counterpart of the  $\gamma$ rays (see, e.g., \cite{Stecker:1978ah}). Indeed, $pp$ collisions produce charged pions, $\pi^\pm$, as efficiently as $\pi^0$, and muon and electron \ns\ ($\nu_\mu, \nu_e$) are then produced by decay, e.g., $\pi^+ \rightarrow \mu^+ + \numu \rightarrow \numu + e^+ + \nue + \barnumu$.  This mechanism is  also the origin of the observed flux of \ns\ from  cosmic-ray interactions in the Earth's atmosphere. 

In this Letter we present the first study of \ns\ from the \fb, in the hadronic model, and show that they are detectable at a kilometer scale experiment, like the existing \ic\ \cite{IceCube}, or the planned \nt\ \cite{KM3NeT}, depending on the detector's location.  This is a new signal, whose detection or exclusion will contribute to understanding the physics of the bubbles. 

\begin{figure}[htbp]
\centering
\includegraphics[width=0.42\textwidth]{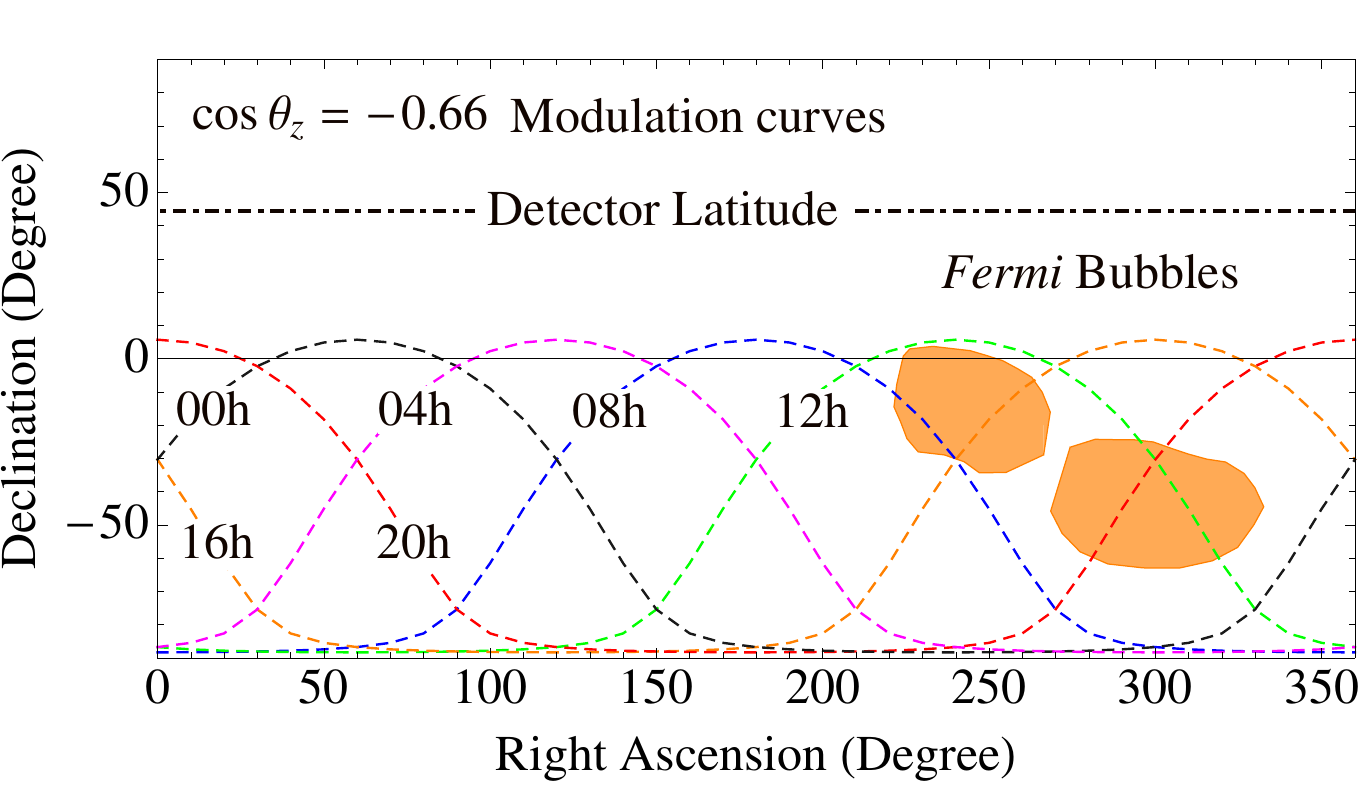}
\caption{The contour of the {\em Fermi} bubbles in equatorial coordinates (shaded), adapted from \cite{Su:2010qj}.  Considering $\theta_z$ as the zenith angle with respect to a neutrino telescope at a latitude $+43^\circ$ (\nt), the area below each curve has $\cos \theta_z <-0.66$ at a given time of the day (labels on curves).}
\label{bubblesfig}
\end{figure}

Figure~\ref{bubblesfig} shows a sketch of the \fb\ in equatorial coordinates. They face the Earth's southern hemisphere, extending to  $\sim 60^\circ$ below the equator, and subtend a solid angle $\Omega_b = 0.808$~sr~\cite{Su:2010qj}. The bubbles are uniformly bright in $\gamma$ rays, with sharp edges and a hard spectrum, $dN_\gamma/dE\propto E^{-2}$~\cite{Su:2010qj}. 

In the hadronic model of Crocker and Aharonian (CA)~\cite{Crocker:2010dg}, the  bubbles are powered by SNRs, which are widely believed to be the factories of cosmic rays up to $10^{15}$--$10^{17}$~eV \cite{SNRs}.  The low-density bubble interiors (average density $\langle n_H \rangle \sim 10^{-2}$~cm$^{-3}$~\cite{Su:2010qj}) are created by  prolonged SF activity near the Galactic center, which forms a high-velocity bipolar wind.   This wind carries the cosmic ray protons produced in  the SNRs to fill the  bubble cavities.   The interactions of these protons with the diluted hot gas produce the \fb\ as observed by  the {\it Fermi}-LAT.  The associated time scale for proton cooling is $\sim 5\times 10^9$~yr~\cite{Crocker:2010dg}, and therefore a SF period of at least this duration is required in the CA model. 

\begin{figure}[htbp]
\centering
\includegraphics[trim = 0.in 0.28in 0.95in 1.in, clip, width=0.42\textwidth]{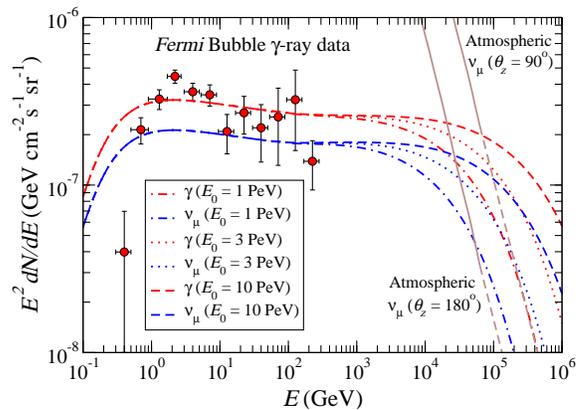}
\caption{The fluxes of $\gamma$ and of $\numu$ (total of \ns\ and antineutrinos)  from the \fb, for different proton cutoff energy $E_0$ (Eq.~(\ref{p-spectrum})), compared with the atmospheric $\nu_\mu$ flux and  the {\it Fermi}-LAT data~\cite{Su:2010qj}.  Errors are $1\sigma$. } 
\label{spectrumfig}
\end{figure}

To calculate the  $\gamma$ and $\nu$ fluxes from $pp$ interactions in the \fb\ we use an injected proton spectrum in the bubble cavities, at steady-state, of the form 
\begin{equation}
N_p(E) = N_0 E^{-k} \exp (-E/E_0),
\label{p-spectrum}
\end{equation} 
where $N_0$, $k$ and $E_0$ are the normalization factor, spectral index and cutoff energy respectively.   The cutoff energy, $E_0 \sim 1-10$ PeV, is  motivated by cosmic-ray acceleration in SNRs~\cite{SNRs}, and $k\sim 2$ is expected in a  Fermi acceleration mechanism.  The index of the injected protons remains hard, in the \fb, due to a saturation condition realized when the acceleration (in the SNRs) time, $pp$ energy loss  (in the FB) time and escape (from the FB) time satisfy a relation:  $t_{acc} < t_{loss} \lesssim t_{esc}$~\cite{Crocker:2010dg}.  Using a parameterization from the Monte Carlo code SIBYLL \cite{Kelner:2006tc}, we obtain the fluxes of gamma rays and neutrinos from $\pi^0$, $\pi^\pm$ decay, above $\gtrsim 10$~GeV.  At lower energies, the fluxes can be modeled with a delta-function approximation. For example, the $\gamma$ flux is given by 
\begin{equation}
\Phi_\gamma (E_\gamma) = \frac{\varphi\langle n_H \rangle}{4\pi D^2 K_\pi} 
\int_{E_{\pi,\rm th}}^\infty dE_\pi\, \frac{\sigma_{pp}(E_c)}
{\sqrt{E_\pi^2 - m_\pi^2}} N_p(E_c) \,.
\label{gamma-delta}
\end{equation}
Here $D$ is the distance to the bubbles, $K_\pi \approx 0.17$ is the mean fraction of proton kinetic energy converting to $\pi^0$, $E_{\pi,\rm th} = E_\gamma +m_\pi^2/4E_\gamma$ is the threshold pion energy, $\sigma_{pp}$ is the inelastic $pp$ cross section, and $E_c = E_\pi/K_\pi + m_p$.  The parameter $\varphi \sim {\cal O}(1)$, and is adjusted to match the $E \gtrsim 10$~GeV Monte Carlo results.  Note that there is a $\sim 10\%$ uncertainty in the hadronic models \cite{Kelner:2006tc}. We also  neglected a contribution of $\sim 10\%$ or less (depending on the unknown magnetic field in the FB) due to Compton-scattered photons off secondary $e^\pm$. Our calculation gives the fluxes of $\gamma$ and of $\numu, \barnumu, \nue, \barnue$, in the expected flavor ratio $\numu\ :   \nue\ = 2 : 1$ for both \ns\ and antineutrinos. Here we focus on $\numu, \barnumu$, for which detectors are the most sensitive.  The fluxes of $\gamma$ and $\numu+ \barnumu$ are shown in Fig.~\ref{spectrumfig}, for $k=2.1$. They are compared to the flux of atmospheric $\numu$, which constitutes the main background.  For this, we used the model in \cite{Honda:1995hz}, extrapolated to fit the high energy \ic\ data \cite{Abbasi:2010ie}. At our energies of interest, flavor oscillations are negligible for atmospheric \ns, due to the short propagation length. However, they are effective for the \fb\ flux, so that the $\numu$ flux at Earth is about 1/2 of the unoscillated one (see e.g.~\cite{Costantini:2004ap}). 

Let us comment on the energetics of the CA model. The total number of $\gamma$-ray producing cosmic-ray protons, in steady state, in the \fb\ is $\int dE~N_p(E) \approx 4.3\times 10^{57}~(\langle n_H \rangle /10^{-2}~{\rm cm}^{-3})^{-1} (D/8.5~{\rm kpc})^2$, where $D=8.5~{\rm kpc}$ is the distance to the bubble centers~\cite{Su:2010qj}.  The total energy in steady state cosmic-ray proton population is $\int dE~N_p(E)E \approx 5.4\times 10^{55}~(\langle n_H \rangle /10^{-2}~{\rm cm}^{-3})^{-1} (D/8.5~{\rm kpc})^2$~ergs.  For comparison, with a $2\times 10^{37}$~erg~s$^{-1}$ power in $\gamma$ rays from both bubbles~\footnote{The power is overestimated by a factor 2 in~\cite{Su:2010qj}, M.~Su, priv.\ commun.} and a $\sim 5\times 10^9$~yr lifetime of the \fb, the total energy emitted in $\gamma$ rays is  $\sim 3\times 10^{54}$~ergs.  This is $\sim 6\%$ of the total cosmic-ray proton energy, an efficiency typical of hadronic models.  

In order to produce the observed uniform surface brightness of FB, a volume emissivity of the form $(R^2-r^2)^{-1/2}; (r< R\approx 3.5$~kpc) seems required~\cite{Su:2010qj}. Whether this can in principle be obtained in the hadronic scenario has not been studied and is beyond our purpose here. A possibility is that anisotropies can arise from the diffusion of cosmic rays in the magnetic field generated by the shocked shells of the bubble~\footnote{R. Crocker, priv.\ commun.}.  In any case, the $\nu$ flux should trace the $\gamma$-ray flux from the \fb, and therefore have the same uniform  projected intensity. 

Under this assumption, we can write the number of $\numu$ events accumulated in a certain exposure time $t_{exp}$, and from a certain zenith bin $[\theta_1, \theta_2]$, (with respect to the detector's position) as follows:
\beq
N_\nu = \int^{t_{exp}}_0  dt  \int_{
        \substack{\Sigma(t) \\ \theta_1 \leq \theta_z \leq \theta_2}
        } d\Omega \int_{E_{th}}^\infty  dE ~\Phi(E) A_\nu(E, \theta_z) 
\nonumber \\
        \simeq t_{exp}  \langle \Omega(\theta_1, \theta_2) \rangle_t    
        \int_{E_{th}}^\infty  dE ~\Phi(E) \langle A_\nu(E)   \rangle_{\theta}~,~~~~~~
\label{nev2}
\eeq
with a completely analogous expression for $\barnumu$.  Here $A_\nu$ is the detector's effective area, $\theta_z$ is the zenith angle with respect to the detector, and $\Sigma(t)$ indicates that the integral in the solid angle is done over the region of the bubbles for which the condition on the zenith angle is satisfied.  This region depends on the time of the day, due to the revolution of the Earth (Fig.~\ref{bubblesfig}).  The approximation in Eq.~(\ref{nev2}) uses the time-averaged solid angle, $\langle \Omega(\theta_1, \theta_2) \rangle_t $, and the effective area averaged over the zenith bin of interest, $\langle A_\nu(E)  \rangle_{\theta}$.  The latter is adequate for sufficiently narrow bins, since $A_\nu$ varies gently with $\theta_z$.

As a benchmark, we use twice  the effective area of the  IC40 configuration of \ic~\cite{Abbasi:2010ie}, $A_\nu = 2 A_{\rm IC40}$, which well represents the full \ic\ configuration~\cite{Montaruli:2011} and is reasonable for \nt.  We  take a 10-year exposure, and consider only the $\numu + \barnumu$ flux in sub-horizon directions, $\theta_z > 90^\circ$, as these channels dominate the effective area. 

\begin{figure}[htbp]
\centering
\includegraphics[width=0.39\textwidth]{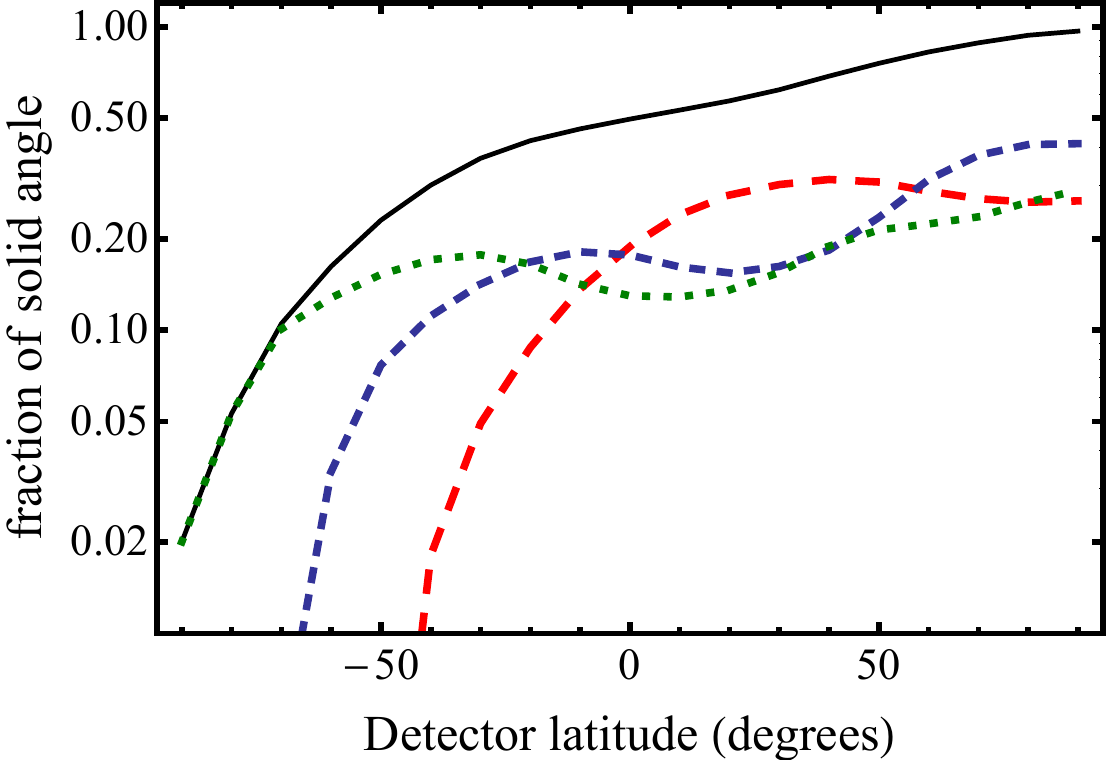}
 \caption{Daily-average fraction of the solid angle subtended by the {\em Fermi} bubbles, $\Omega_b=0.808$~sr, that falls in a given bin of zenith angle, $\theta_z$ (with respect to the detector), as a function of the detector latitude.  The bins in $\cos\theta_z$ are: $[-1, -0.66]$ (long dashed), $[-0.66,-0.33]$ (short dashed), $[-0.33, 0]$ (dotted) and [-1,0] (total sub-horizon, solid). }
\label{solidangle}
\end{figure}

Under this condition, a detector in the northern hemisphere is more sensitive to the  \fb\ compared to a southern one since the Galactic center is in the southern hemisphere (Fig.~\ref{bubblesfig}). This appears clearly in Fig.~\ref{solidangle}, which shows the ratio $\langle \Omega(\theta_1, \theta_2) \rangle_t /\Omega_b$, for different zenith bins of interest, and as a function of the detector's latitude, $\theta_{lat}$.  IceCube, at $\theta_{lat}=-90^\circ$, is sensitive only to  the small portion of the upper bubble that lies in the northern hemisphere (Fig.~\ref{bubblesfig}), for which $\langle \Omega \rangle_t \approx 0.02$~sr.  For KM3NeT we used $\theta_{lat}=+43^\circ$, the position of the current ANTARES detector in the Mediterranean~\cite{Collaboration:2011nsa}, and we have a much more encouraging total of $\langle \Omega(\cos \theta_z <0) \rangle_t  \approx 0.57$~sr (Fig.~\ref{solidangle}). Note that this location is nearly optimal (maximum solid angle) for events in the deepest bin, $\cos \theta_z < -0.66$, where the signal to background ratio is the highest (see below). This bin corresponds to a cone of opening angle $\sim 48^\circ$, which is wide enough to contain both bubbles almost completely when the center of the lower bubble is at $\theta_z \sim 180^\circ$ from the detector (Fig.~\ref{bubblesfig}).     

\begin{figure}[htbp]
\centering
\includegraphics[trim = 0.15in 0.25in 1.in 1.in, clip, width=0.39\textwidth]{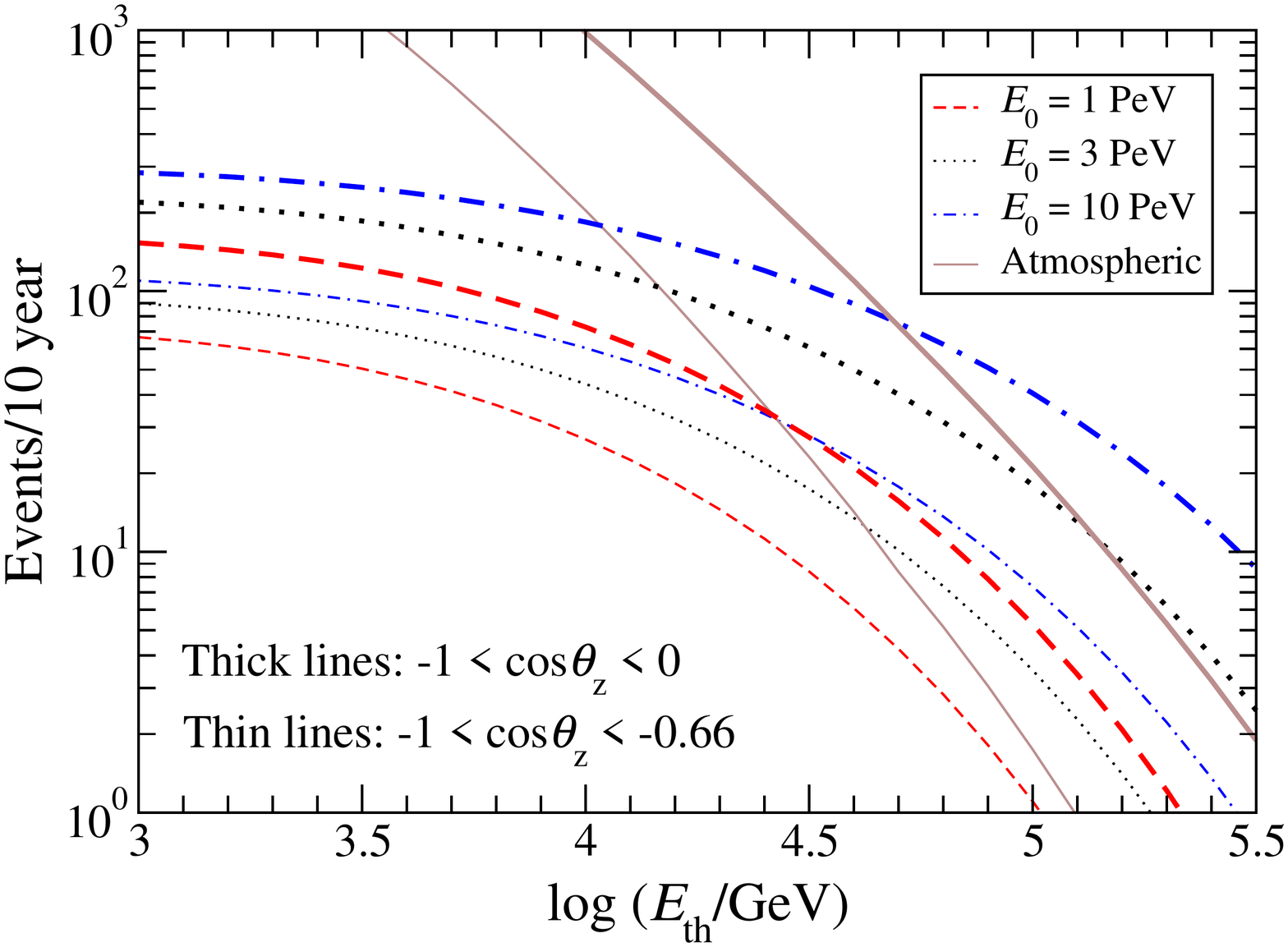}
\includegraphics[width=0.39\textwidth]{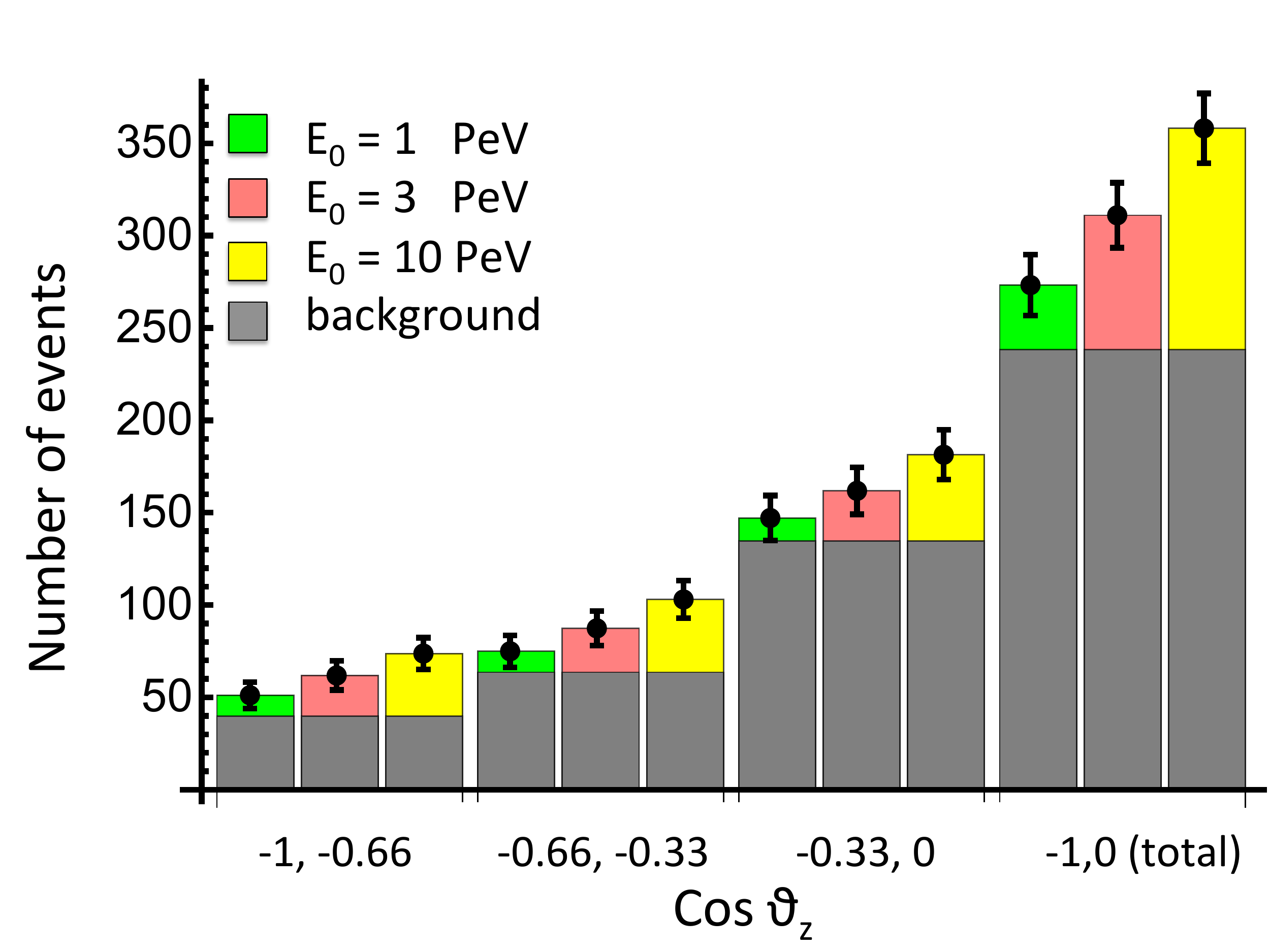}
\caption{ Number of signal and background events in 10 years, at a km scale detector at $+43^\circ$ latitude, for different cutoff energy $E_0$ in the proton spectrum.  Top: as a function of the energy threshold, $E_{th}$.  Bottom: zenith distribution, for $E_{th}=25$ TeV.  Expected $1\sigma$ error bars are shown.}
\label{eventrate}
\end{figure}

Figure~\ref{eventrate} (top panel) shows the number of signal and background events at \nt, as a function of $E_{th}$ for the entire region of sensitivity ($\cos\theta_z < 0$) and for the deepest zenith bin ($\cos\theta_z < -0.66$).  They are in rough agreement with the earlier estimates in \cite{Crocker:2010dg}.  

Counting on the good  directional sensitivity of the detector, we restrict the analysis of the background to the  daily-averaged solid angles subtended by the bubbles, for better signal discrimination.  For the entire sub-horizon solid angle (solid line in Fig.~\ref{solidangle}), and for $E_0=3, 10$ PeV, the signal exceeds the background above $\sim 50-100$ TeV, where $\sim 20-100$ signal events are expected.  We found that $E_{th} \sim 20-30$ TeV gives the maximum statistical excess of the signal over the background, $\Delta = N_{sig}/\sqrt{N_{sig } + N_{bkg}}$, which is between 4 and 6 $\sigma$ (i.e., $\Delta \sim 4-6$).  

For the deepest zenith bin the event rate falls faster with $E_{th}$, compared to the entire region of zenith, reflecting the stronger opacity of the Earth to $\nu$'s crossing its diameter. However the atmospheric flux is also suppressed due to the strongest absorption of muons.  Therefore the signal is higher than background already above $\sim 20$ TeV in the most optimistic case.  The statistical significance of the signal is lower due to the lower number of events, but it exceeds $3 \sigma$ for $E_0 = 3, 10$ PeV, nevertheless. 

The zenith distributions of the signal and background are given in Fig.~\ref{eventrate} (bottom panel), together with the quantity  $\sqrt{N_{sig } + N_{bkg}}$ in the form of error bars.  In all cases, the numbers of signal events in the three bins are similar.  This is because the strongest \n\ absorption in the deepest bin (lower $A_\nu$) compensates for the larger solid angle.  The figure also confirms how the number of events in the deepest bin and in the total sub-horizon region are the best indicators of the effect,  in terms of statistical significance, while the other two bins individually suffer from low statistics and high background. 

Some remarks are in order.  We have taken $k=2.1$, but results obtained with a much more conservative cosmic-ray proton spectrum, $k=2.3$, also fit the $\gamma$-ray data \cite{Crocker:2010dg}.  In this case, the $\nu$ event rate is dominated by background, and $\Delta \sim 1.2 \sqrt{t_{exp}/10~{\rm yr}}$, for $\cos\theta_z < 0$.  A $\sim 6$ times longer exposure would be required to reach $3 \sigma$.  Note, however, that using $\Delta > 3$ as criterion of statistical significance is conservative.  Indeed, an excess below 3$\sigma$ in number of events is likely to be further substantiated, in a full fit of the data, by other indicators like the correlation with the position and extent of the \fb.  

Briefly, let us comment on the sensitivity of a south pole detector. Due to the suppression in the solid angle and the near-horizon direction of the signal, the number of signal events is low and the background level is enhanced compared to \nt.  For the best case scenario of $E_0=10$ PeV, we find $N_{sig}=8.0$ and $N_{bkg}=59$ for $E_{th}=10$ TeV, and  $N_{sig}=2.4$ and $N_{bkg}=1.7$ for $E_{th}=100$ TeV.  Thus, an excess above background will not be statistically significant, unless an exposure at least $\sim 6$--9 times larger is achieved  or efforts to gain sensitivity to events above the horizon \cite{Abbasi:2009cv} and to better discriminate background (e.g., by detecting tau neutrinos \cite{icetau}) are successful.  While a discovery with \ic\ seems unlikely, still an analysis of the \ic\ data might place interesting constraints on the parameters: for example, it may be able to exclude larger, less natural, values of $E_0$ or harder proton spectra, $k<2.1$.

In summary, the \fb\ are a potential new source of galactic high energy \ns.  For spectral cutoff $E_0 \sim 1$--10~PeV, and spectral index $k \sim 2.1$ for cosmic-ray protons -- motivated by particle acceleration in SNRs --  the \n\ flux could exceed the atmospheric background above $\sim 20-50$ TeV and be seen at km-scale detectors.  For a steeper spectrum, detection will require longer exposures or larger detectors.  \ic\ is strongly disadvantaged by its southern location, but it could obtain important constraints, and might achieve detection with an increased sensitivity to the sky above the horizon.  Our results motivate efforts in this direction.  For a northern detector, observing the \fb\ with a decade of operation is a realistic goal, that contributes to make the case for the planned \nt\ in the Mediterranean.  It is encouraging that the Italian node of \nt\ is funded~\cite{infn}. 

The observation of \ns\ from the \fb\ would strongly support the hadronic origin of $\gamma$ rays from these mysterious objects and favor the model for $\sim 10^9$ years scale activity at the Galactic center to form the \fb\ \cite{Crocker:2010dg}.  The shape of the \n\ spectrum would help to distinguish this scenario from more exotic \n-emitting explanations of the \fb, like dark matter annihilation or decay~\cite{Malyshev:2010xc}.  A null result would indicate a softer proton spectra or a non-hadronic mechanism as the primary origin of $\gamma$ rays and favor $\sim 10^6$ years scale activity by Sgr A$^\star$ to form the \fb\ \cite{Su:2010qj}.

As \n\ and  $\gamma$-ray detectors progress, it is likely that their interplay will ultimately identify the nature and origin of the \fb. Implications will be far reaching, showing what high energy phenomena are at play in our galaxy and, by extension, in other galaxies too.

\acknowledgements 
We thank F.~Aharonian, R.~Crocker, N.~Kurahashi and M.~Su for useful discussions.  Supports from the NSF Grant No.\ PHY-0854827 (CL) and NASA Fermi Cycle 4 Guest Investigator Program NNH10ZDA001N (SR) are acknowledged.  Work of SR was performed at and while under contract with the U.S.~Naval Research Laboratory.

\newpage

{\large \bf Supplementary material}

\setcounter{figure}{0}
\renewcommand{\thefigure}{S-\arabic{figure}}
\renewcommand{\thetable}{S-\Roman{table}}

\begin{figure}[htbp]
\centering
\includegraphics[trim = 0.in 0.28in 0.95in 1.in, clip, width=0.42\textwidth]{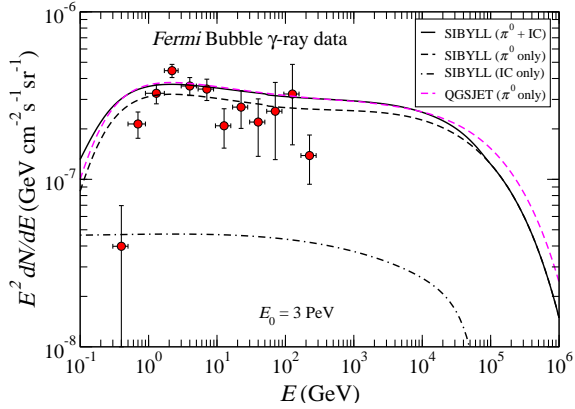}
\caption{Contribution to the observed $\gamma$-ray spectrum from Compton-scattered CMB photons, labeled SIBYLL (IC only), by $\pi^\pm$ decay $e^\pm$ in case the bubble magnetic field is $\ll 3$~$\mu$G.  The IC component, the normalization of which depends on the unknown bubble magnetic field, is ignored in Fig.~\ref{spectrumfig} of the main text.  Also shown is the $\gamma$-ray spectrum from another hadronic model, labeled QGSJET ($\pi^0$ only), for the same injected proton spectrum, to illustrate hadronic model uncertainties.}
\label{spectrumR1}
\end{figure}

\begin{figure}[htbp]
\centering
\includegraphics[trim = 0.in 0.28in 0.95in 1.in, clip, width=0.42\textwidth]{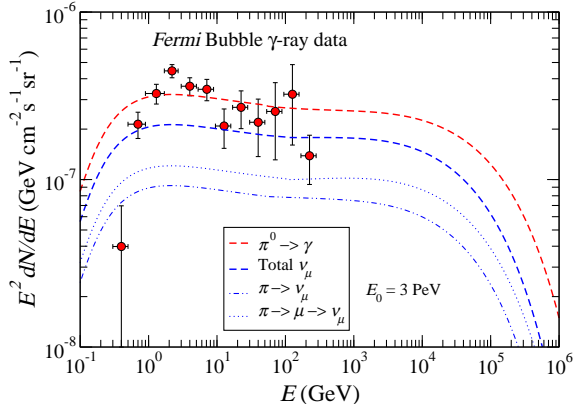}
\caption{Neutrino and antineutrino flux components before oscillation from $\pi^\pm$ chain decay for the total $\nu_\mu + {\bar \nu}_\mu$ flux, labeled as Total $\nu_\mu$, plotted here and in Fig.~\ref{spectrumfig} of the main text.  The $\nu_e + {\bar \nu}_e$ component is very similar to the $\pi \to \mu \to \nu_\mu$ component plotted here.}
\label{spectrumR2}
\end{figure}

\begin{figure}[htbp]
\centering
\includegraphics[trim = 0.3in 0.35in 1.25in 1.in, clip, width=0.48\textwidth]{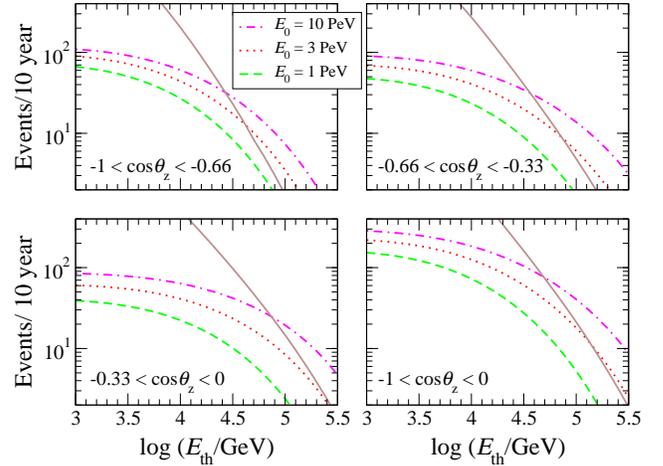}
\caption{Number of signal and background events in 10 years, for 3 zenith-angle bins and the total, in a km scale detector at +43 degree latitude (KM3NeT). The events are calculated as a function of the neutrino threshold energy and for up to 1 PeV energy. The event rates in the top left and bottom right panels are reported in Fig.~\ref{eventrate} of the main text.}
\label{evts4panel}
\end{figure}

\begin{figure}[htbp]
\centering
\includegraphics[trim = 0.3in 0.35in 1.25in 1.in, clip, width=0.48\textwidth]{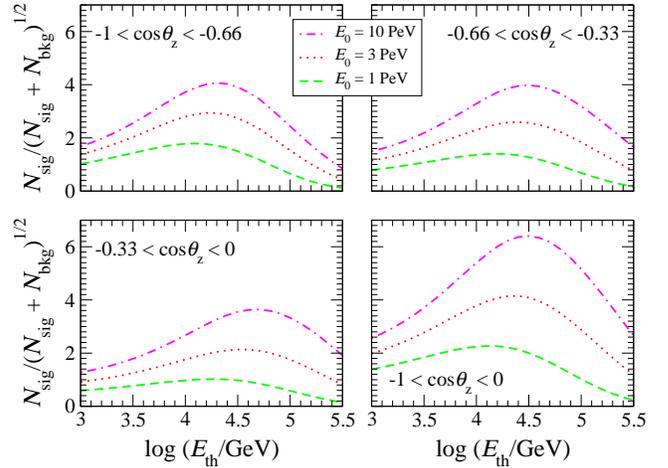}
\caption{Statistical significance of the bubble neutrino signal for the events in Fig.~\ref{evts4panel} above.}
\label{sig4panels}
\end{figure}

\newpage

\begin{table}[h]
\caption{\label{tab1} Number of signal and background events for a km scale detector at the South pole (IceCube). The daily-averaged solid angle is 0.02 sr (see Fig.~\ref{solidangle} in the main text).}
\begin{ruledtabular}
\begin{tabular}{lcc}
Events/10~yr & $E_{\rm th} =10^4$~GeV & $E_{\rm th} = 10^5$~GeV \\
\hline 
Signal ($E_0 = 1$~PeV) & 2.8 & 0.3 \\
Signal ($E_0 = 3$~PeV) & 5.1 & 1.1 \\
Signal ($E_0 = 10$~PeV) & 8.0 & 2.4 \\
Background & 59 & 1.7
\end{tabular}
\end{ruledtabular}
\end{table}

\begin{figure}[htbp]
\centering
\includegraphics[trim = 0.in 0.28in 0.95in 1.in, clip, width=0.42\textwidth]{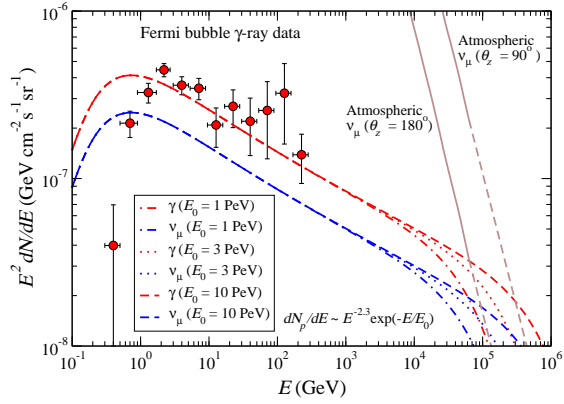}
\caption{Gamma-ray and neutrino flux models from the Fermi bubbles in case of a conservative cosmic-ray proton spectral index of $-2.3$, instead of $-2.1$ used in Fig.~\ref{spectrumfig} of the main text and all previous analyses.}
\label{spectrumfig2.3}
\end{figure}

\begin{figure}[htbp]
\centering
\includegraphics[trim = 0.25in 0.05in 0.95in 0.1in, clip, width=0.42\textwidth]{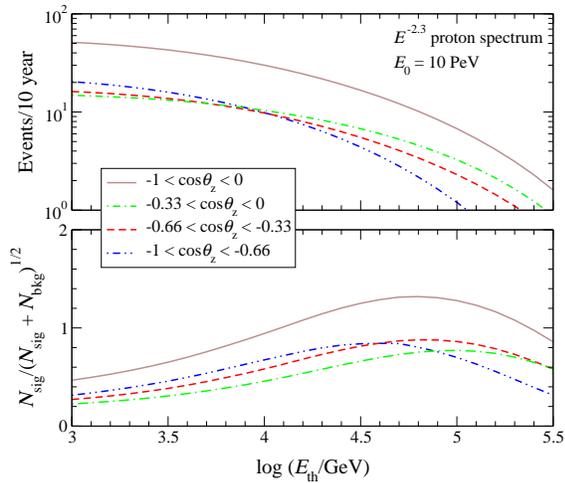}
\caption{Detection prospects of neutrinos from the Fermi bubbles for the cosmic-ray proton spectral index of $-2.3$ (Fig.~\ref{spectrumfig2.3}), at a km scale detector at +43 degree latitude (KM3NeT). Top -- Number of signal and background events in 10 years. Bottom -- Detection significance for 10 years}
\label{evtdetfig2.3}
\end{figure}

\end{document}